\documentclass[10pt]{article}
\pdfoutput=1
\usepackage{amsmath,amssymb}
\usepackage{amsthm}
\usepackage{mathptmx}
\usepackage{cite}
\usepackage{latexsym}
\numberwithin{equation}{section}

\usepackage{hyperref}

\begin{document}

\title{Buchdahl type inequalities in $d$-dimensions}

\author{
Matthew Wright\footnote{matthew.wright.13@ucl.ac.uk}\\
Department of Mathematics, University College London\\
Gower Street, London, WC1E 6BT, UK
}

\date{\today}

\maketitle

\begin{abstract}
Spherically symmetric anisotropic static compact solutions to the Einstein equations in dimension $d\geq4$ are considered. Various matter models are examined and upper bounds on the ratio of the gravitational mass to the radius in these different models are obtained, and saturation of these bounds are proven. Bounds are also generalised in the presence of a non-zero charge and a positive cosmological constant. These bounds are then used to find the maximum of the gravitational redshift at the surface of the object.  
\end{abstract}

\section{Introduction}

Determining bounds on the mass-radius ratio for spherically symmetric spacetimes is an important question; it allows one to calculate the maximum degree of compactification of a body and to obtain an upper bound on the gravitational redshift of the object. The well known Buchdahl theorem~\cite{Buchdahl:1959zz} tells us that when we have a spherically symmetric isotropic object whose energy density is non-increasing outwards, then the mass-radius ratio must satisfy the bound
\begin{align}
\frac{2M}{R}\leq \frac {8}{9}, \label{Buchdahl}
\end{align}
where $M$ is the ADM mass of the asymptotically flat spacetime and $R$ is the radius of the body in Schwarzschild coordinates, defined by the location of the surface of vanishing pressure. This bound has a number of important implications. It tells us that the corresponding Schwarzschild radius of a fluid always lies inside the sphere, and thus a regular matter distribution cannot describe the interior of a black hole. The existence of such a bound is also of interest as it occurs well before the appearance of an apparent horizon at $M=R/2$.

However, as pointed out in~\cite{Guven:1999wm}, the assumptions in Buchdahl's theorem are quite restrictive; there are many physical systems that do not satisfy them. For example a soap bubble does not obey the non-increasing energy density condition. Moreover, many models of stars do not satisfy the isotropy condition. Thus obtaining bounds on the mass radius ratio while removing Buchdahl's assumptions is an important question. This was considered in~\cite{Baumgarte:1993,Mars:1996gd} where they were able to show under very general conditions the important result that 
\begin{align}
\frac{2M}{R}<1.
\end{align}
However obtaining a stronger bound than this is of great interest, since for example, this bound alone still allows the possibility of an unbounded gravitational redshift.

Andr\`{e}asson was able to derive bounds on the mass radius without assuming the Buchdahl assumptions~\cite{Andreasson:2007ck}; and considered the class of models satisfying the energy condition
\begin{align}
p+2p_{\bot}\leq \Omega\rho, \label{EVcondition}
\end{align}
where $p$ is the radial pressure, $p_{\bot}$ is the tangential pressure, $\rho$ is the energy density and $\Omega$ is a constant which determines the matter model. Additionally, all these quantities are assumed to be non-negative. This energy condition is very general. For example the dominant energy condition tells us that the energy density must be greater than the pressure in any direction. Thus setting $\Omega=3$, and assuming the radial and tangential pressures are positive, gives us a condition which must be satisfied by matter obeying the dominant energy condition. Whereas letting $\Omega=1$ gives a condition which must be satisfied by the spherically symmetric Einstein-Vlasov system, a system which admits a very wide class of physically important solutions; the overwhelming majority of which do not obey either of the Buchdahl assumptions. Assuming this energy condition, the bound
\begin{align}
\frac{2M}{R}\leq\frac{(1+2\Omega)^2-1}{(1+2\Omega)^2}
\end{align}
was obtained, which coincides with Buchdahl's inequality~(\ref{Buchdahl}) in the case $\Omega=1$.

Further bounds were derived in~\cite{Karageorgis:2007cy}, where different restrictions on the energy and pressure are considered; we review these bounds in Section 3. Buchdahl's theorem has been generalised to the case of a positive cosmological constant in~\cite{BalagueraAntolinez:2004sg,Boehmer:2003uz,Boehmer:2005sm,Boehmer:2006ye,Mak:2001gg}, and Buchdahl's assumptions were removed in~\cite{Andreasson:2009pe}. Bounds have also been derived assuming a non-zero charge, initially using Buchdahl's assumptions, see for example~\cite{HaMa00}, and again Andr\`{e}asson~\cite{Andreasson:2008xw} removed these assumptions and instead assumed condition~(\ref{EVcondition}) with $\Omega=1$. Finally bounds on $M/R$ were derived in the presence of both a positive cosmological constant and non-zero charge in~\cite{Andreasson:2012dj}.
 
In this paper we consider $d$-dimensional spherically symmetric spacetimes, where $d\geq4$. The natural quantity to consider in $d$-dimensions is the ratio 
\begin{align}
\frac{M}{R^{d-3}}.
\end{align}
This is the natural ratio to consider as it is the quantity appearing in the higher dimensional Schwarzschild metric and thus it allows one to compute the upper bound on the gravitational redshift of the body.

Studying higher dimensional compact objects in general relativity has interested many authors. The higher dimensional Buchdahl inequality for a perfect fluid was derived in~\cite{PoncedeLeon:2000pj}, and is given by
\begin{align}
\frac{\tilde{M}}{R^{d-3}}\leq\frac{2(d-2)}{(d-1)^2}, \label{Buchdahld}
\end{align}
where $\tilde{M}$ is related to the mass of the fluid. In~\cite{Zarro:2009gd} this bound is generalised to include a non-zero cosmological constant, generalising the inequality found in~\cite{Mak:2001gg}. An inequality for $d$-dimensional charged spheres was also derived in~\cite{HaMa00}.  The equivalent of the Buchdahl bound has also been considered in five dimensional Gauss-Bonnet gravity~\cite{Wright:2015yda}. However all the proofs of these inequalities involved assuming Buchdahl's assumptions. In this paper we derive various bounds on the ratio $M/R^{d-3}$ in higher dimensional general relativity, removing Buchdahl's restrictive assumptions and instead considering the various general matter models considered in~\cite{Andreasson:2007ck,Karageorgis:2007cy}. We also find bounds involving both a positive cosmological constant and a non-zero charge. 

This paper is organised as follows. In Section 2 we set-up Einstein's equations and define our mass functions. In Section 3 we discuss the different matter models we consider and review previous known bounds in four dimension. In Section 4 we state our main results both with and without a cosmological constant and charge.  In Section 5 we provide the proof of these inequalities, and then in Section 6 we discuss the saturation of these bounds.

\section{Equations for $d$-dimensional anisotropic fluid sphere}
We will consider a spherically symmetric anisotropic matter distribution; and we begin by working in full generality with both a non-zero charge and cosmological constant. In the interior we will assume spherically symmetric anisotropic charged matter, with energy density $\rho$, radial pressure $p$, tangential pressure $p_{\bot}$ and proper charge density $\sigma$. The Einstein Maxwell equations with cosmological term are
\begin{align}
G_{ij}+\Lambda g_{ij}=\kappa\left( T^{({\rm m})}_{ij}+T^{({\rm EM} )}_{ij} \right),
\\ \partial_{j}(\sqrt{-g}F^{ij})=\sqrt{-g}J^{i}, \quad \partial_{[i}F_{jk]}=0,
\end{align}
where $\kappa=8\pi G_d$ is the $d$-dimensional gravitational coupling constant.

We write the metric in the spherically symmetric form
\begin{align}
ds^2=-e^{a(r)}dt^2+e^{b(r)}dr^2+r^2 d\Omega^2_{d-2}.
\end{align}
We a priori do not assume these coordinates cover the whole of our space time, we begin simply assume they cover a patch around the origin. They will only be valid up to the first sphere in which $e^{-b(r)}=0$. It will be a conclusion of our results that in fact we can choose the coordinates to be global, and in that case if the matter source is compact we can then match to an exterior solution. With these coordinate the energy momentum tensor of matter $T^{({\rm m})}_{ij}$ is given by the $d \times d$ matrix
\begin{align}
T^{({\rm m})}{}^{i}_{j}={\rm diag}(-\rho,p,p_{\bot},p_{\bot},....,p_{\bot}).
\end{align}

The energy momentum tensor of the electromagnetic field is given by
\begin{align}
T^{({\rm EM} )}_{ij}=F_{ik}F_{j}{}^{k}-\frac{1}{d}g_{ij}F^{mn}F_{mn} .
\end{align}
The $d$-current density vector is given by
\begin{align}
J^{i}=A_{d-2} \sigma u^{i}.
\end{align} 
Here $u^{i}$ is the $d$-velocity satisfying $u^iu_i=-1$ and $A_{d-2}$ is the area of the unit $(d-2)$-sphere given by
\begin{align}
A_{d-2}=\frac{2\pi^{\frac{d-1}{2}}}{\Gamma(\frac{d-1}{2})},
\end{align}
where $\Gamma$ is the gamma function.

Let us define the charge within a $(d-2)$ sphere of radius $r$ given by
\begin{align}
q(r)=A_{d-2}\int_{0}^{r}e^{(a+b)(r')}r'^{d-2}\sigma d r'.
\end{align}
We will also assume that the electromagnetic distribution has identically zero magnetic part. We can now solve Maxwell's equations; and we find the only non-zero component of the electromagnetic field tensor to be 
\begin{align}
F_{tr}=\frac{e^{(a+b)/2}}{r^{d-2}}q.
\end{align}
Now, using this solution for the field tensor, we find that in the interior the Einstein field equations become
\begin{align}
&\kappa  \rho+\Lambda+\frac{\kappa q^2}{2A_{d-2}r^{2(d-2)}}=\frac{(d-2)}{2r^{d-2}}\frac{d}{dr}\left(r^{d-3}(1-e^{-b})\right), \label{feq1}
\\
&\kappa p -\Lambda-\frac{\kappa q^2}{2A_{d-2}r^{2(d-2)}}=\frac{(d-2)e^{-b}}{2}\left[ \frac{ a'}{r}+\frac{(d-3)}{r^2}\right]-\frac{(d-3)(d-2)}{2r^2}, \label{feq2}
\\
&\kappa p_{\bot}-\frac{\kappa q^2}{2A_{d-2}r^{2(d-2)}}-\Lambda=\label{feq3}\\&\frac{e^{-b}}{2}\left[a''+\frac{a'^2}{2}-\frac{a'b'}{2}-\frac{(d-3)(b'-a')}{r}+\frac{(d-3)(d-4)}{r^2}\right]-\frac{(d-3)(d-4)}{2r^2}.\nonumber
\end{align}

The first of these equations,~(\ref{feq1}) can be integrated immediately to find
\begin{align}
e^{-b}=1-\frac{2\kappa m(r)}{(d-2)A_{d-2}r^{d-3}}-\frac{\kappa f(r)}{(d-2)A_{d-2}r^{d-3}}-\frac{2\Lambda r^2}{(d-2)(d-3)}
\end{align}
where we have defined the function $f$ by
\begin{align}
f(r):=\int_{0}^{r}\frac{q(r')^2}{r'^{d-2}}dr',
\end{align}
and the mass function $m$ \footnote{To compare our results with those derived in~\cite{PoncedeLeon:2000pj}, we should note that they use the normalisation for the mass function
\begin{align}
m(r)=\frac{\kappa}{d-2}\int_{0}^{r}\rho(r')r'^{d-2}dr'.
\end{align}
Our definition agrees with that given in~\cite{Myers:1986un}, which we prefer as evaluating at $r=R$ in the absence of $\Lambda$ and charge gives us the physical ADM mass of the spacetime. In the bound~(\ref{Buchdahld}) derived in~\cite{PoncedeLeon:2000pj}, the constant $\tilde{M}$ is related to the ADM mass $M$ by the relation
\begin{align}
M=\frac{(d-2)A_{d-2}}{\kappa}\tilde{M},
\end{align}
which in four dimensions simply becomes $\tilde{M}=GM$.} 
as
\begin{align}
m(r):=\int \rho \, dV =A_{d-2}\int_{0}^{r}\rho(r')r'^{d-2}dr',
\end{align}
where the last equality simply follows from spherical symmetry. In four dimensions this corresponds to the Newtonian mass as no metric terms are included.  

If the source is compact, then the exterior solution is found by setting the energy density and pressure to zero in the Einstein field equations~(\ref{EFE1})-(\ref{EFE3}) and is given by $d$-dimensional Reissner-Nordstr\"{o}m-de Sitter solution; first derived in~\cite{Myers:1986un,Xu:1988ju}
\begin{align}
e^{a(r)}=e^{-b(r)}=1-\frac{2\mu}{r^{d-3}}+\frac{\Theta^2}{r^{2d-6}}-\frac{2\Lambda r^2}{(d-1)(d-2)},
\end{align}
where $\Lambda$ is the cosmological constant. The constants $\mu$ and $\Theta$ are related to the total mass and charge; $M$ and $Q$ respectively, by
\begin{align}
\mu&=\frac{\kappa}{(d-2)A_{d-2}}M,\\
\Theta^2&=\frac{\kappa}{(d-2)(d-3)A_{d-2}}Q^2.
\end{align}

Now assuming the interior solution is of finite extent we denote its boundary by the radius $r=R$. Requiring that the interior and exterior solutions match on the boundary of this surface $r=R$ tells us that the total mass
\begin{align}
M=m(R)+\frac{Q^2}{2(d-3)R^{d-3}}+\frac{f(R)}{2}.
\end{align}
Thus we can define the interior gravitational mass function $m_g=m_g(r)$ as
\begin{align}
m_g(r)=m(r)+\frac{q(r)^2}{2(d-3)r^{d-3}}+\frac{f(r)}{2},
\end{align}
which evaluating at the boundary gives
\begin{align}
M=m_g(R).
\end{align}
When the electromagnetic charge vanishes, the mass functions $m$ and $m_g$ coincide.

\section{Matter models and previous results}
 In this section we will discuss the various matter models that we will consider, and review previous bounds that have been obtained. All results stated here will assume vanishing charge and $\Lambda=0$ unless explicitly stated otherwise. We will consider the following distinct matter models:
\begin{itemize}
\item {\bf Buchdahl's assumptions } Assuming the energy is non-increasing outwards gives rise to the famous Buchdahl bound
\begin{align}
\frac{2M}{R}\leq \frac{8}{9}.
\end{align} 
This was generalised to $d$-dimensions in~\cite{PoncedeLeon:2000pj} where is was found (using our notation)
\begin{align}
\frac{2\kappa m_g}{(d-2)A_{d-2}r^{d-3}}\leq \frac{4(d-2)}{(d-1)^2}.
\end{align}
In~\cite{Zarro:2009gd} this is generalised to include a non-zero cosmological constant where it is found (rewritten into our notation)
\begin{align}
\frac{\kappa M}{(d-2)A_{d-2}R^{d-3}}\leq \left[1-\frac{2\Lambda R^2}{(d-2)(d-1)}\right]\left[ 1-\frac{1}{(d-1)^2}\frac{\left(d-3-\frac{\Lambda}{\bar{\rho_f}})^2\right)}{(1-\frac{2\Lambda R^2}{(d-2)(d-1)})}\right],
\end{align} 
where $\bar{\rho}_f:=\frac{(d-1)M}{A_{d-2} R^{d-1}}$ is the mean fluid density. We will not consider Buchdahl's assumptions further in this paper. 
\item {\bf Dominant radial pressure} $p_{\bot}\leq p$. In particular, this bound encompasses the important sub-case of a perfect fluid; when we have isotropic pressure $p_{\bot}=p$. In four dimensions the bound
\begin{align}
\frac{2m_g}{r}\leq 12\sqrt{2}-16 \approx 0.9705 \label{4disotropic}
\end{align}
was first derived by Bondi in~\cite{Bondi:1964zz} and the proof was made rigorous in~\cite{Karageorgis:2007cy}. This inequality is less strict than the Buchdahl bound~(\ref{Buchdahl}).
\item {\bf Dominant energy in the tangential direction} $p_{\bot}\leq \rho$. For most reasonable matter models, from the dominant energy condition one expects $|p_{\bot}|\leq \rho$, and assuming positivity of tangential pressure this is equivalent to $p_{\bot}<\rho$. Assuming simply $p_{\bot}\leq \rho$, without the assumption of positivity of $p_\bot$, the sharp bound
\begin{align}
\frac{2m_g}{r}\leq \frac{2\sqrt{2}+2}{5}\approx 0.9657 \label{4ddet}
\end{align} 
was derived in~\cite{Karageorgis:2007cy}. Again this bound is weaker than the Buchdahl bound.
\item{\bf Andr\`{e}asson condition} The condition $p+(d-2)p_{\bot}\leq \,  \Omega \rho$
was first studied in~\cite{Andreasson:2007ck}. This inequality is very general and incorporates a number of realistic fluid models. For example when $\Omega=d-1$, any solution with positive energy and pressure obeying the dominant energy condition is admitted as a solution, whereas $\Omega=1$ corresponds to assuming the trace of the energy momentum tensor is negative. In four dimensions this is a condition which must be satisfied by Einstein-Vlasov matter, and this inequality is the natural generalisation of this condition to consider in higher dimensions. The bound derived in~\cite{Andreasson:2007ck} was
\begin{align}
\frac{2m_g}{r}\leq\frac{(1+2\Omega)^2-1}{(1+2\Omega)^2}. \label{4dAnder}
\end{align}
We present a simplified proof of this inequality in this paper, inspired by the method used in~\cite{Karageorgis:2007cy}, and generalise it to $d$-dimensions. 
Bounds have also been derived in the presence of a non-zero cosmological constant and charge; assuming this condition with $\Omega=1$. In~\cite{Andreasson:2012dj} it was found that assuming the condition
\begin{align}
0\leq\frac{q^2}{r^2}+\Lambda r^2 \leq 1,
\end{align}
then the following bound can be derived
\begin{align}
\frac{m_g}{r}\leq \frac{2}{9}+\frac{q^2}{3r^2}-\frac{\Lambda r^2}{3}+\frac{2}{9}\sqrt{1+\frac{3q^2}{r^2}+3\Lambda r^2}. \label{4dEVcharge}
\end{align}
We generalise this bound to higher dimensions.
\item {\bf Zero radial pressure with  Andr\`{e}asson condition} $p_{\bot}\geq p=0$ A solution describing a zero radial pressure solution was considered by Florides in~\cite{Florides:1974}, where the bound
\begin{align}
\frac{2m_g}{r}\leq \frac{2}{3}
\end{align} 
was derived assuming the  Andr\`{e}asson condition with $\Omega=1$. Such a model can be useful for describing infinitely thin shells like soap bubbles, which will have no radial pressure. We will briefly examine such a situation here.
\end{itemize}

We are now in a position to formulate our main results.

\section{Main results}

\subsection{Inequalities without charge and cosmological constant}
Consider a $d$-dimensional space-time  solution to Einstein's equations and assume the pressures $p$, $p_{\bot}$ and energy density $\rho$ are all non-negative. Then we have the following bounds on the ratio of gravitational mass $m_g$ to the radius $r^{d-3}$:
\begin{enumerate}
\item Dominant radial pressure, $p_{\bot}\leq p$.
\begin{align}
 \frac{2\kappa m_g}{(d-2)A_{d-2}r^{d-3}}\leq\frac{4(d-1)\sqrt{d-2}-8(d-2)}{(d-3)^2}.\label{isotropic}
\end{align}
\item Dominant energy in tangental direction, $p_{\bot} \leq \rho$. 
\begin{align}
\frac{2\kappa m_g}{(d-2)A_{d-2}r^{d-3}}\leq \frac{(4 d-12) \sqrt{(d-3)(d-2)}+2(d-2)(2d-7)}{(3
   d-10) (3 d-7)}. \label{dominant}
\end{align}
\item  Andr\`{e}asson condition, $p+(d-2)p_{\bot}\leq \Omega \rho$.
\begin{align}
\frac{2\kappa m_g}{(d-2)A_{d-2}r^{d-3}}\leq \frac{(d-3+2\Omega)^2-(d-3)^2}{(d-3+2\Omega)^2}.\label{EV}
\end{align}
In the particular case of $\Omega=1$ with zero radial pressure, this bound is improved to
\begin{align}
\frac{2\kappa m_g}{(d-2)A_{d-2}r^{d-3}}\leq \frac{2}{d-1}. \label{Florides1}
\end{align}
\end{enumerate}

\subsection{Inequalities with charge and $\Lambda$ }
We can also readily generalise some of these inequalities to include both charge and a cosmological constant. Here we only consider the case of dominant energy in the tangential direction and the  Andr\`{e}asson condition with $\Omega=1$.
Consider any solution to the Einstein-Maxwell equations with $\Lambda\geq0$ and charge $q(r)$ satisfying the condition 
\begin{align}
\frac{\kappa q(r)^2}{(d-2)(d-3)A_{d-2}r^{2d-6}}+\frac{2\Lambda r^2}{(d-2)(d-3)}\leq 1 . \label{assumption1}
\end{align}
Then we have the following inequalities:
\begin{enumerate}
\item Dominant energy in tangential direction, $p_{\bot} \leq \rho$.
For $d=4$:
\begin{align}
\frac{2m_g}{r}\leq \frac{1}{5}\left( 2+\frac{ 4q^2}{r^{2}}-\frac{8\Lambda r^2}{3} +2\sqrt{2(1+\frac{ q^2}{r^{2}}+\Lambda r^2 )}\right). \label{dominant2}
\end{align}
For $d\geq5$:
\begin{multline}
\frac{2\kappa m_g(r)}{(d-2)A_{d-2}r^{d-3}}\leq\frac{2\kappa  q(r)^2}{(3d-7)(d-3)A_{d-2}r^{2d-6}}-\frac{8\Lambda r^2}{(d-1)(3d-7)} \\+\frac{1}{(3
   d-10) (3 d-7)}(2(d-2)(2d-7)\\+(4 d-12) \sqrt{(d-3)(d-2)+(3 d-10) (\frac{\kappa  q(r)^2}{(d-2)A_{d-2}r^{2d-6}}+\frac{2\Lambda r^2}{(d-2)})}). \label{dominant3}
\end{multline}

\item   Andr\`{e}asson condition with $\Omega=1$,  $p+(d-2)p_{\bot}\leq\rho$. 
\begin{multline}
\frac{\kappa  m_g}{(d-2)A_{d-2}r^{d-3}}\leq \frac{(d-2)}{(d-1)^2}\\+\frac{\kappa  q^2}{(d-1)(d-2)(d-3)A_{d-2}r^{2d-6}}-\frac{2\Lambda r^2}{(d-1)(d-2)}\\+\frac{(d-1)}{(d-2)^2}\sqrt{1+(d-1)\left(\frac{\kappa  q^2}{(d-2)A_{d-2}r^{2d-6}}
+\frac{2\Lambda r^2}{(d-2)}  \right)} .\label{EV2}
\end{multline}
\end{enumerate}

\subsection{Discussion of results}

Let us make a few remarks about these inequalities. To compare these results to the four dimensional counterpart, let us note that the setting $d=4$ on the left hand side of these inequalities gives
\begin{align}
\frac{2\kappa m_g}{(d-2) A_{d-2}r^{d-3}}=\frac{2 G \,  m_g}{r}.
\end{align}
Thus we observe that the inequalities~(\ref{isotropic}), (\ref{EV}), (\ref{dominant}) all reduce to the four dimensional inequalities~(\ref{4disotropic}), (\ref{4dAnder}) respectively which were derived in~\cite{Karageorgis:2007cy}. Like in the four dimensional case, the bound~(\ref{EV}) also reduces to the $d$-dimensional Buchdahl inequality~(\ref{Buchdahld}) when one sets $\Omega=1$.  

In four dimensions, assuming the pressure in the radial direction is greater than the tangential pressure gives a stricter bound compared to assuming that the energy is greater than the tangential pressure.  In five dimensions and higher this remains true. 

Assuming positive pressure, the dominant energy condition tells us that both $p_{\bot}\leq \rho$ and $p\leq \rho$. Thus the inequality $p+(d-2)p_{\bot}\leq (d-1)\rho$ holds, which is equivalent to the Andr\`{e}asson condition  with $\Omega=d-1$. Inserting this into~(\ref{EV}) yields

\begin{align}
\frac{2\kappa m_g}{(d-2)A_{d-2}r^{d-3}}\leq \frac{8(d-1)(d-2)}{(3d-5)^2}.\label{DEC}
\end{align}
However assuming $p_{\bot}\leq \rho$ only gives the bound~(\ref{dominant}), which must also be satisfied by matter with positive pressure obeying the dominant energy condition. This inequality~(\ref{dominant}) is in fact stronger than the above inequality~(\ref{DEC}) in all dimensions greater than four.

Inequalities~(\ref{dominant2}) and (\ref{EV2}) generalise two of these inequalities to include both charge and a positive cosmological constant, assuming the inequality~(\ref{assumption1}) holds. In $d=4$ this assumption reduces to that considered in~\cite{Andreasson:2012dj}, and our condition is the natural generalisation of this to higher dimensions. To the best of the authors knowledge, the $d=4$ inequality~(\ref{dominant2}) has not been previously derived, and is displayed separately to highlight this fact. The inequality for the  Andr\`{e}asson condition with $\Omega=1$~(\ref{EV2}) reduces to the result found in~\cite{Andreasson:2012dj}.

Evaluating these inequalities at the boundary $r=R$ will give us bounds on the ratio $M/R^{d-3}$. In the absence of charge and cosmological constant, the ratio
\begin{align}
\frac{2\kappa M}{(d-2)A_{d-2}R^{d-3}}
\end{align}
appearing in the $d$-dimensional Schwarzschild metric will give us bounds on the gravitational redshift at the surface of the star. In particular the redshift $z$ at the surface will be
\begin{align}
z=\left(1-\frac{2\kappa M}{(d-2)A_{d-2}R^{d-3}}\right)^{-\frac{1}{2}}-1.
\end{align}
Thus, assuming our different energy conditions, we obtain bounds on $z$, which we display in Table~\ref{tab01}.
\begin{table}[!ht]
\begin{center}
\begingroup
\renewcommand*{\arraystretch}{1.8}
\begin{tabular}{|c|c|} \hline
Energy condition & Upper bound on $z$  \\
\hline \hline
Buchdahl &  $\frac{2}{d-3}$ \\ \hline
$p_{\bot} \leq p$ & $\frac{d-3}{\sqrt{d^2+2d-7-4(d-1)\sqrt{d-2}}}-1 $ \\ \hline
$p_{\bot}\leq \rho$ &  $\frac{ \sqrt{(3d-7)(3d-10)}}{\sqrt{d-3} \sqrt{5 d-14-4 \sqrt{(d-3)(d-2)}}}-1 $ \\ \hline
$p+(d-2)p_{\bot}\leq \Omega\rho$ &  $\frac{2\Omega}{d-3}$ \\ \hline 
$p=0$, $(d-2)p_{\bot}\leq\rho$ &  $\sqrt{\frac{d-1}{d-3}}$ \\ \hline
\end{tabular}
\endgroup
\caption{Bounds on $z$ for different matter models.} 
\label{tab01}
\end{center}
\end{table} 

This allows us to ask the question in which dimension is the gravitational redshift greatest. The right hand side of the above upper bounds are all maximal in $d=4$ and all decay monotonically with increasing dimension.  Likewise examining the $g_{00}$ component of the Schwarzschild metric, we will see that it is closer to unity in higher dimensions. These observations agree with the conclusion in~\cite{PoncedeLeon:2000pj} that the effects of gravity are greatest in four dimensions. 

We should note that the strongest of the inequalities (ignoring the zero radial pressure case, which is a very strict assumption) in Table~\ref{tab01} is given by assuming the condition $p+(d-2)p_{\bot}\leq \Omega\rho$ with $\Omega\leq 1$. This is also true in four dimensions, however this inequality decays much faster in higher dimensions than the respective inequalities for dominant radial pressure or dominant energy in the tangential direction.

\section{Proof of Bounds}
Inspired by the proofs considered in~\cite{Karageorgis:2007cy,Andreasson:2012dj} for the four dimensional inequalities; let us introduce new variables $x$, $y$, $z_1$ and $z_2$ defined as follows
\begin{align}
x&=\frac{2\kappa m_g(r)}{(d-2)A_{d-2}r^{d-3}}-\frac{\kappa  q(r)^2}{(d-2)(d-3)A_{d-2}r^{2d-6}}+\frac{2\Lambda r^2}{(d-1)(d-2)}, \label{x}
\\y&=\frac{2\kappa }{d-2} r^2 p, \label{y}
\\z_1&=\frac{\kappa  q(r)^2}{(d-2)A_{d-2}r^{2d-6}},
\\z_2&=\frac{2\Lambda r^2}{(d-2)}.  
\end{align}
Requiring that the metric is to be regular everywhere, along with assumption~(\ref{assumption1}), allows us to deduce these variable lie in the set
\begin{align}
\mathcal{U}_1: =\{(x,y,z_1,z_2) \in  \mathbb{R}^4 \, | \, 0\leq x<1, y\geq 0, z_1\geq0, z_2\geq0, z_1+z_2 \leq (d-3) \}.
\end{align}
In particular the assumption~(\ref{assumption1}) ensures the condition $z_1+z_2\leq d-3$ holds, which is necessary for what follows. Let us also introduce the new variable $\beta=2\log r$. Then it can easily be shown that the Einstein field equations~(\ref{feq1})-(\ref{feq3}) can be rewritten in terms of these new variables as
\begin{align}
	\frac{2\kappa }{(d-2)}\rho r^2&=2\dot{x}+(d-3)x-z_1-z_2, \label{EFE1}
	\\ \frac{2\kappa }{(d-2)} pr^2 &=y, \label{EFE2}
	\\ 	2\kappa  p_{\bot}r^2 &= \frac{((d-3)x+y-z_1-z_2)\dot{x}}{(1-x)}\nonumber\\&+\frac{((d-3) x+y-z_1-z_2)^2}{2 (1-x)}+2\dot{y}-2\dot{z_1}+2z_1-y(4-d), \label{EFE3}
\end{align}
where $\dot{x}=\frac{dx}{d\beta}$.

\subsection{Dominant radial pressure case}
Let us first consider the case of dominant radial pressure case, which also includes the important sub-case of isotropic pressure. We will only consider the case without charge and $\Lambda$, so we simply set $z_1$ and $z_2$ to zero in the Einstein field equations. This condition $p\geq p_{\bot}$ can be rewritten in terms of our new variables $x$ and $y$ as
\begin{multline}
((d-3)x+y)\dot{x}+2(1-x)\dot{y}\leq 2y(1-x)-\frac{((d-3 ) x+y)^2}{2 }
=: -v_1(x,y).
\end{multline}
Now let us define the function
\begin{align}
w_1=\frac{((d-3)(1-x)+(d-3)+y)^2}{1-x}.
\end{align}
Then differentiating $w_1$ with respect to $\beta$ we find
\begin{multline}
\dot{w_1}=\frac{((d-3)(1-x)+(d-3)+y)}{(1-x)^2}((d-3)x+y)\dot{x}+2(1-x)(\dot{y})\\\leq-\frac{((d-3)(1-x)+(d-3)+y)}{(1-x)^2} v_1(x,y). \label{w1dot}
\end{multline}
By assumption the variables $x$ and $y$ lie in the set $\mathcal{U}_1$, and hence we can deduce 
\begin{align}
((d-3)(1-x)+(d-3)+y)>0.
\end{align}
Thus $w_1$ is decreasing whenever $v_1>0$. Hence the supremum of $w_1$ occurs when $v_1\leq0$ and this is readily seen to be
\begin{align}
w_1\leq \sup_{\mathcal{U}_1, v_1\leq0} w_1=w_1(0,4)=4(d-1)^2.
\end{align}
And thus setting $y=0$ we can show the maximum of $x$ in this range occurs at
\begin{align}
x=\frac{4(d-1)\sqrt{d-2}-8(d-2)}{(d-3)^2},
\end{align}
which after reinserting the definition for $x$ proves inequality~(\ref{isotropic}).

\subsection{Dominant energy in the tangential direction}
Now let us assume that the energy density is greater than the tangential pressure $p_{\bot}\leq\rho$, and this time we will not set $z_1$ or $z_2$ equal to zero.  Using~(\ref{EFE1})-(\ref{EFE3}) we find
\begin{multline}
((3d-7)x+y-z_1-z_2-2(d-2))\dot{x}+2(1-x)(\dot{y}-\dot{z_1}-\dot{z_2})\\\leq(1-x)((d-2)(d-3)x-(d-4)y-2z_1-2z_2)-\frac{1}{2}((d-3)x+y-z_1-z_2)^2
\\ =:-v_2(x,y,z_1,z_2) .
\end{multline}
This time we define the function $w_2$ as
\begin{align}
w_2(x,y,z_1,z_2):= \frac{((3d-7)(1-x)+(d-3)+y-z_1-z_2)^2}{1-x}.
\end{align}
And now differentiating with respect to $\beta$ we find
\begin{multline}
\dot{w_2}= \frac{((3d-7)(1-x)+(d-3)+y-z_1-z_2)}{(1-x)^2}\\ \times((3d-7)x+y-z_1-z_2-2(d-2))\dot{x}+2(1-x)(\dot{y}-\dot{z_1}-\dot{z_2})
\\ \leq -\frac{((3d-7)(1-x)+(d-3)+y-z_1-z_2)}{(1-x)^2} v_2 \label{w_2dot}
\end{multline}
Now, since the variables lie in the set $\mathcal{U}_1$, we have that
\begin{align}
 ((3d-7)(1-x)+(d-3)+y-z_1-z_2)\geq0,
\end{align}
and so $w_2$ is decreasing whenever $v_2\geq0$. Thus
\begin{align}
w_2\leq \sup_{\mathcal{U}_1, v_2\leq0} w_2.
\end{align}
 Now in the case $d=4$ the supremum of $w_2$ occurs at $(x,y,z_1,z_2)=(1/10,1/2,0,0)$, in which case $w_2=40$. Hence we can conclude
\begin{align}
x\leq \frac{1}{5}(2-z_1-z_2+2\sqrt{2(1+z_1+z_2)}),
\end{align}
and thus~(\ref{dominant2}) immediately follows. For $d\geq5$ we find the supremum of $w_2$ occurs at
\begin{align}
\left(\frac{3d-11}{(3d-7)(3d-10)},\frac{d-3}{3d-10},0,0\right)
\end{align}
with 
\begin{align}
\sup_{\mathcal{U}_1, v_2\leq0} w_2= \frac{16(d-3)^2(3d-7)}{3d-10}
\end{align}
And hence setting $y=0$ we find that $x$ must satisfy
\begin{align}
x\leq \frac{(4 d-12) \sqrt{(d-3)(d-2)+(3 d-10) (z_1+z_2)}-(3 d-10) (z_1+z_2)+2(d-2)(2d-7))}{(3
   d-10) (3 d-7)},
\end{align}
which after reinserting the definitions of $x, y, z_1, z_2$ we find the inequality~(\ref{dominant3}). Setting the charge $q$ and cosmological constant $\Lambda$ equal to zero, we immediately recover~(\ref{dominant}).

\subsection{Andr\`{e}asson condition}
We now consider the condition 
\begin{align}
p+(d-2)p_{\bot}\leq \Omega\rho,
\end{align}
in the absence of charge and cosmological constant. Rewriting this condition in terms of $x$ and $y$ we find
\begin{multline}
\dot{x}((d-3+2\Omega)x+y-2\Omega)+2\dot{y}(1-x)
\\\leq -\frac{1}{2} \left( ((d-3)+2\Omega)(d-3)x^2+y^2-2\Omega(d-3) x+2(d-3)y \right)
\\=:-\frac{1}{2}v_3(x,y).
\end{multline}
This time we define
\begin{align}
w_3=\frac{((d-3+2\Omega)(1-x)+y+d-3)^2}{1-x},
\end{align}
and again we differentiate this with respect to $\beta$
\begin{multline}
\dot{w_3}=\frac{((d-3+2\Omega)(1-x)+y+d-3)}{(1-x)^2}(\dot{x}((d-3+2\Omega)x+y-2\Omega)+2\dot{y}(1-x))
\\
\leq-\frac{((d-3+2\Omega)(1-x)+y+d-3)}{2(1-x)^2} v_3(x,y). \label{w3dot}
\end{multline}
As previously if $v_3\geq 0 $ then $w$ is decreasing so the supremum of $w_3$ satisfies
\begin{align}
w_3\leq \sup_{\mathcal{U}_1, v_3\leq0} w_3=w_3(0,0)=4(d-3+\Omega)^2.
\end{align}
Thus, setting $y=0$ we derive
\begin{align}
x\leq \frac{(d-3+2\Omega)^2-(d-3)^2}{(d-3+2\Omega)^2},
\end{align}
and reinserting the definition of $x$ yields the bound~(\ref{EV}).

\subsection{Andr\`{e}asson condition with $\Omega=1$ in presence of $\Lambda$ and $q$}
Now we restrict our attention to the energy condition associated with the trace of the energy momentum tensor being negative
\begin{align}
p+(d-2)p_{\bot}\leq \rho,
\end{align}
and reintroduce charge and $\Lambda$ into our equations. Rewriting this in terms of $x$, $y$, $z_1$ and $z_2$ gives
\begin{multline}
y+\frac{((d-3)x+y-z_1-z_2)\dot{x}}{(1-x)}+\frac{((d-3) x+y-z_1-z_2)^2}{2 (1-x)}\\+2\dot{y}-2\dot{z_1}+2z_1-y(4-d)
\leq 2\dot{x}+(d-3)x-z_1-z_2.
\end{multline}
Rearranging this gives us
\begin{multline}
\dot{x}((d-1)x+y-z_1-z_2-2)+2(\dot{y}-\dot{z_1}-z_2)(1-x)
\\\leq -\frac{1}{2} \left( (11-6d+d^2)x^2+(y-z_1-z_2)^2-2((d-3)x-(d-3)y)-2(dx-3)(z_1+z_2) \right)
\\=:-\frac{1}{2}v_4(x,y,z_1,z_2).
\end{multline}
As before, let us define
\begin{align}
w_4=\frac{(2(d-2)-(d-1)x+y-z_1-z_2)^2}{1-x},
\end{align}
and differentiating $w_4$ gives us
\begin{multline}
\dot{w_4}=\frac{(2(d-2)-(d-1)x+y-z_1-z_2)}{(1-x)^2}(\dot{x}((d-1)x+y-z_1-z_2-2)\\+2(\dot{y}-\dot{z_1}-z_2)(1-x)) \leq -\frac{1}{2}\frac{(2(d-2)-(d-1)x+y-z_1-z_2)}{(1-x)^2}v_4(x,y,z_1,z_2).
\end{multline}
Now, $2(d-2)-(d-1)x+y-z_1-z_2\geq0$ since the variables lie in the set $\mathcal{U}_1$. If $v_4\geq 0 $, then $w_4$ is decreasing so
we can hence deduce that 
\begin{align}
w_4\leq\sup_{\mathcal{U}_1, v_4\leq0}w_4=w_4(0,0,0,0)=4(d-2)^2.
\end{align}
Now taking $y=0$ this implies
\begin{align}
x\leq \frac{1}{(d-1)^2}\left( 2(d-2) -(d-1)(z_1+z_2)+2(d-2)\sqrt{1+(d-1)(z_1+z_2)}\right).
\end{align}
If we now substitute the expressions for $x$, $z_1$ and $z_2$ back into this, we find the inequality~(\ref{EV2}).
\subsection{Zero-radial pressure case with $\Omega=1$  Andr\`{e}asson condition}
Finally we consider one last sub-case. Setting the radial pressure equal to zero, we can derive the bound~(\ref{Florides1}). Let us set the charge equal to zero and assume the  Andr\`{e}asson condition with $\Omega=1$. We find
\begin{align}
\frac{((d-3)x-z_2)\dot{x}}{(1-x)}+\frac{((d-3) x-z_2)^2}{2 (1-x)}\leq2\dot{x}+(d-3)x-z_2. \label{zepre}
\end{align}
Since the energy $\rho>0$, we can divide by the right hand side of this inequality to find
\begin{align}
\frac{(d-3)x-z_2}{2(1-x)}\leq 1,
\end{align}
which rearranging gives
\begin{align}
x\leq\frac{2+z_2}{d-1},
\end{align}
which after reinserting the definitions of $x$ and $z_2$ gives the inequality~(\ref*{Florides1}). We should note that this proof works for any non-zero cosmological constant, and the bound is independent of this cosmological constant.

\section{Sharpness of bounds}

In this section we investigate the sharpness of the derived bounds in the absence of charge and a cosmological constant. Let us briefly review the issue of sharpness of the corresponding four dimensional bounds. For the cases of dominant radial pressure and dominant energy in the tangential direction, without charge and a cosmological constant, the saturation of the four dimensional bounds~(\ref{4ddet}) and~(\ref{4disotropic}) was shown in~\cite{Karageorgis:2007cy}. For the Einstein-Vlasov system, it was shown in~\cite{Andreasson:2006ab}~that an infinitely thin shell solution uniquely saturates the bound~(\ref{4dAnder}) with $\Omega=1$, and~\cite{Karageorgis:2007cy} later also showed sharpness of this bound employing a different technique. In the case where charge is included into this Einstein-Vlasov system, the bound~(\ref{4dEVcharge}) (with $\Lambda=0$) was shown to be sharp in~\cite{Andreasson:2008xw}, however a numerical investigation performed in~\cite{Andreasson:2009qu} indicates that this saturating solution may not be unique. As soon as a cosmological constant is included into the system the question of sharpness remains an open problem~\cite{Andreasson:2009pe}, with infinitely thin shell solutions shown not to generically saturate the inequalities. 

In this paper to examine the question of sharpness we will generalise the methods used in~\cite{Karageorgis:2007cy}. We will see that the generalised higher dimensional bounds of Section 4.1 do remain saturated in higher dimensions. We will not investigate further the sharpness of the bounds in the presence of charge and cosmological constant. The methods of~\cite{Karageorgis:2007cy} do not readily generalise to these cases, since phase space is now four dimensional and so constructing a saturating solution is a more difficult problem.  

The structure of the proof is as follows. We continue to work with the variables $x$ and $y$ as defined in~(\ref{x}),(\ref{y}), and we construct a curve in the $(x,y)$ plane which is arbitrarily close to the point $(x_B,0)$, where $x_B$ is the appropriate bound obtained on $x$ in the three matter models considered in section 4.1. We then use this curve to construct a solution to Einstein's equation, with this solution obeying the appropriate energy condition. 

All variables defined below retain their same definition as in the previous section. We begin the proof by constructing a parametric curve
\begin{align}
x=x(\tau), \quad y=y(\tau); \quad \tau \in (0,\infty)
\end{align}
with the following properties
\begin{enumerate}
\item $\frac{1}{\nu_i} \frac{dw_i}{d\tau}$ for $i=1,2,3$ is both  negative and is integrable
\item $0\leq x(\tau)<1, \,\, y(\tau)\geq0 $ for all $\tau$.
\item There exists $\tau_0$ such that $y(\tau)=0$ for all $\tau>\tau_0$ and $x(\tau)\rightarrow 0$ as $\tau\rightarrow \infty$.
\item The curve is $\mathcal{C}^1$ except at a finite number of points.
\end{enumerate}
were the case $i=1$ corresponds to the dominant radial pressure case, $i=2$ corresponds to the dominant energy in the tangential direction, and $i=3$ corresponds to the case of the Andr\`{e}asson condition. The third condition is the requirement that our solution be asymptotically flat, and in fact we will see  that all of our saturating solutions are compact. 

Once we have constructed such a curve, we can then construct the solution which saturates the bound. To do this we define
\begin{align}
\kappa_1(\tau)&=-\frac{1}{\nu_1}\frac{dw_1}{d\tau}\frac{(1-x)^2}{(d-3)(2-x)+y}
\\
\kappa_2(\tau)&=-\frac{1}{\nu_2}\frac{dw_2}{d\tau}\frac{(1-x)^2}{(3d-7)(1-x)+(d-3)+y}
\\
\kappa_3(\tau)&=-\frac{1}{\nu_3}\frac{dw_3}{d\tau}\frac{2(1-x)^2}{2(d-3+\Omega)-(d-3+2\Omega)x+y}
\end{align}
with the conditions (1) and (2) listed above telling us that $\kappa_i$ is positive and integrable. 

Now we can construct our solution as follows, for $i=1,2,3$ we let
\begin{align}
\beta=\int \kappa_i \, d\tau, \quad r=e^\frac{\beta}{2}.
\end{align}
The definitions of the $\kappa_i$ will now ensure that
\begin{align}
\dot{w_i}=\frac{1}{\kappa_i} \frac{dw_i}{d\tau}
\end{align}
will give equality in the inequalities found for $\dot{w_i}$ in equations~(\ref{w1dot}),(\ref{w_2dot}),(\ref{w3dot}). This also implies the energy conditions are all saturated. It can easily be seen using~(\ref{feq2}) and~(\ref{EFE2}) that the following relation holds between the metric coefficient $a$ and the new variables $x$ and $y$
\begin{align}
\frac{da}{d\beta}=\frac{(d-3)x+y}{1-x}
\end{align}
This allows us to construct the metric corresponding to the saturating solution as so 
\begin{align}
b(r)=-\log(1-x), \quad a(r)= \int \frac{(d-3)x+y}{2(1-x)}\kappa_i \, d\tau .
\end{align}

\subsection{Dominant radial pressure case}
Now to show the estimate~(\ref{dominant}) is sharp, so we need to construct a spacetime by defining a curve in $(x,y)$ coordinates that intersects a small neighbourhood of the point
\begin{align}
(x,y)=\left(\frac{4(d-1)\sqrt{d-2}-8(d-2)}{(d-3)^2},0\right) \label{pointDR}
\end{align}
while satisfying the four conditions mentioned above.

Let us fix a sufficiently small $\epsilon>0$ and define
\begin{align}
x_\epsilon=\epsilon , \quad y_\epsilon =2-(d-1)\epsilon+2\sqrt{(1-\epsilon)(1-(d-2)\epsilon)},
\end{align}
so that $(x_\epsilon,y_\epsilon)$ lies on the curve $\nu_1=0$. This point is arbitrarily close to the point $(0,4)$ where the maximum of $w_1$ is obtained. The point~(\ref{pointDR}) lies on the curve $w_1(x,y)=4(d-1)^2$. We construct a solution, a portion of which is arbitrarily close to this curve.

The first part of the curve we define implicitly by
\begin{align}
\sqrt{w_1(x,y)}=2(d-3)+2 \epsilon \delta_\epsilon x- \delta_\epsilon x^2
\end{align}
where $\delta_\epsilon$ is defined so that the curve passes through the point $(x_\epsilon,y_\epsilon)$
\begin{align}
\delta_\epsilon=\frac{1}{\epsilon^2}(\sqrt{w_1(x_\epsilon,y_\epsilon)}-2(d-3))
\end{align}
We travel along this curve from the origin to the point $(x_\epsilon,y_\epsilon)$. After that the curve
\begin{align}
\sqrt{w_1(x,y)}=\sqrt{w_1(x_\epsilon,y_\epsilon)}-\frac{\epsilon(x-x_\epsilon)^2}{\sqrt{1-x}} \label{C12}
\end{align}
is followed from the point $(x_\epsilon,y_\epsilon)$ in the direction of increasing $x$ until the $x$-axis is reached. The curve then travels along the $x$-axis until we arrive back at the origin. Let us denote the curve obtained in such a manner $C_1$. Now it is easy to see $C_1$ satisfies the latter three required conditions. It just remains to show the first condition holds along all parts of the curve.

Now along the first part of the curve it can be shown that $\nu_1<0$. And thus since
\begin{align}
\frac{1}{2w_1}\frac{dw_1}{d\tau}=2\delta_\epsilon (x_\epsilon-x) \frac{dx}{d\tau}
\end{align}
$x$ is increasing and $x\leq x_\epsilon$ we require only that $\delta_\epsilon$ is positive. This is true for sufficiently small $\epsilon$ since $\lim\limits_{\epsilon\rightarrow0} \epsilon^2\delta_\epsilon=4$. 

For the second part of $C_1$ defined by~(\ref{C12}), we have that $\nu_1>0$ in this region and $x>x_\epsilon$ and so we just need to calculate
\begin{align}
\frac{1}{2\sqrt{w_1}}\frac{dw_1}{d\tau}=-\epsilon (x-x_\epsilon)\frac{4(1-x)+x-x_\epsilon}{2(1-x)^{\frac{3}{2}}} \frac{dx}{d\tau}<0.
\end{align}
Finally for the part of $C_1$ lying along the $x$-axis, since after setting $y=0$ we find the dimensionally independent result
\begin{align}
\frac{1}{\nu_1}\frac{dw_1}{d\tau} =\frac{2(2-x)}{x(1-x)^2}\frac{dx}{d\tau}<0
\end{align}
where the inequality follows as we are travelling along the direction of decreasing $x$. It is readily checked that the energy and pressure of this solution are everywhere non-negative throughout this curve,  hence we have constructed a curve with the required properties. 

\subsection{Dominant energy in the tangential direction}
This time we need to construct a curve satisfying
\begin{align}
\frac{1}{\nu_2} \frac{d w_2}{d \tau}<0
\end{align}
We construct the curve, which we will denote by $C_2$ with the tangential pressure equal to the energy $p_{\bot}=\rho$. The initial part of the curve is given implicitly by the equation
\begin{align}
\sqrt{w_2(x,y)}=4d-10+\frac{Ax}{3d-7}-Ax^2
\end{align}
$A$ is chosen so that the curve passes through the point where the supremum of $w_2$ occurs, which we denote by
\begin{align}
(x_M,y_M)=\left(\frac{3d-11}{(3d-7)(3d-10)},\frac{d-3}{3d-10}\right). \label{pointDE}
\end{align}
This tells us
\begin{align}
A=\frac{2 \left(2 (d-3)\sqrt{\frac{3d-7}{3 d-10}}-2 d+5\right) \left(9
   d^2-51 d+70\right)^2}{3 d-11}>0\
\end{align}
We follow this curve from the origin to this point~(\ref{pointDE}), and then we follow the curve
\begin{align}
\sqrt{w_2(x,y)}=\sqrt{w_2(x_M,y_M)}-\frac{\epsilon(x-x_M)^2}{\sqrt{1-x}} \label{C22}
\end{align}
in the direction of increasing $x$ until we reach the $x$-axis. Once there we again stay on the $x$-axis and travel in the direction of decreasing $x$ until we reach the origin. This curve is very similar to the curve considered in the previous case, and it is very similar to check that the first condition holds, so we omit the details here.

\subsection{Andr\`{e}asson condition}
This time we will construct a curve which passes arbitrarily close to the point 
\begin{align}
(x,y)=\left(\frac{(d-3+2\Omega)^2-(d-3)^2}{(d-3+2\Omega)^2},0\right).  \label{pointAnd}
\end{align}
Again, we will construct a parametric curve $(x(\tau),y(\tau))$, this time with the property holding
\begin{align}
\frac{1}{\nu_3} \frac{d w_3}{d\tau} <0.
\end{align}

We define the following function
\begin{align}
W_\epsilon (x,y)=\frac{((d-3+2\Omega)(1-\epsilon)(1-x)+y+d-3)^2}{1-x}.
\end{align}
Let us define a sufficiently small $\epsilon>0$ and consider the curve given implicitly by
\begin{align}
W_\epsilon(x,y)=\left( \epsilon\sqrt{(d-3)(1+(d-3+2\Omega)x)}+2(d-3+\Omega)(1-\epsilon)\right)^2 \label{c3}
\end{align} 
When $\epsilon=0$ this reduces to the curve $w_3=4(d-3+\Omega)^2$, which passes through the origin and the desired point~(\ref{pointAnd}) and when $\epsilon=1$ this is simply the curve $\nu_3=0$, which passes through the origin and the point $(\frac{2\Omega}{d-3+2\Omega},0)$. For values of $\epsilon$ between $0$ and $1$ the curve lies between these two curves. We define the curve $C_3$ to be the curve obtained by first travelling from the origin to the point on the $x$-axes, and then travel along the $x$-axis back to the origin. 

Again the latter three required properties are easily shown to be satisfied on this curve by construction. It just remains for us to examine the first property. Let us first consider the part of the curve defined by~(\ref{c3}). Now since $\epsilon=1$ corresponds to the curve $\nu_3=0$, it is straightforward to observe that $\nu_3>0$ along this part of the curve for $\epsilon<1$. And thus we must check whether $\frac{d w_3}{d\tau}<0$. We start by differentiating~(\ref{c3}) to find
\begin{align}
\frac{dW_\epsilon}{d\tau}&=\frac{(d-3+2\Omega)(d-3)\epsilon\sqrt{W_\epsilon}}{\sqrt{(d-3)(d-3+(d-3+2\Omega)x)}}\frac{dx}{d\tau}
\\&=\frac{(d-3+2\Omega)(d-3)\epsilon}{\sqrt{(d-3)(d-3+(d-3+2\Omega)x)}}\frac{((d-3+2\Omega)(1-\epsilon)(1-x)+y+d-3)}{\sqrt{1-x}}\frac{dx}{d\tau}.
\end{align}
However, differentiating the definition of $W_\epsilon$ directly gives
\begin{align}
\frac{dW_\epsilon}{d\tau}=\frac{\left( ((d-3+2\Omega)(1-\epsilon)(1-x)+y+d-3)\right)}{(1-x)^2}\times\\ ((d-3-(d-3+2\Omega)(1-\epsilon)(1-x)+y) \frac{dx}{d\tau}+2(1-x)\frac{dy}{d\tau})
\end{align}
Comparing these two expressions allows us to solve $\frac{dy}{d\tau}$
\begin{multline}
2(1-x)\frac{dy}{d\tau}=\nonumber\\\left( \frac{(d-3+2\Omega)(d-3)^{\frac{1}{2}}(1-x)^{\frac{3}{2}}\epsilon}{\sqrt{(d-3)+(d-3+2\Omega)x}}+(d-3+2\Omega)(1-\epsilon)(1-x)-y-(d-3)\right)\frac{dx}{d\tau}.
\end{multline}
Now we use this to calculate $\frac{dw_3}{d\tau}$ using~(\ref{w3dot})
\begin{align}
\frac{dw_3}{d\tau}= \frac{(d-3+2\Omega)\epsilon}{1-x}((d-3+2\Omega(1-x)+y+d-3)) \left( \sqrt{\frac{(d-3)(1-x)  }{d-3+(d-3+2\Omega)x}}-1\right),
\end{align}
which is negative since the term in the rightmost bracket is negative. And hence we conclude
\begin{align}
\frac{1}{\nu_3}\frac{d w_3}{d\tau}<0
\end{align}
along this part of the curve. For the remaining part of the curve lying along the $x$-axis, we see that by differentiating $w_3$ and setting $y=0$
\begin{align}
\frac{1}{\nu_3}\frac{d w_3}{d\tau}=\frac{1}{(d-3)x}\frac{2(d-3+\Omega)-(d-3+2\Omega)x}{(1-x)^2}\frac{dx}{d\tau}<0
\end{align}
since $\frac{dx}{d\tau}<0$ along this part of the curve.

\subsection{Zero pressure case}
This case is significantly simpler to handle than the previous three cases since our space is now only one dimensional. This time the saturating solution is simply a constant density solution. If one chooses a density 
\begin{align}
\rho_0=\frac{(d-1)(d-2)(d-3)}{\kappa},
\end{align}
then one finds up to the boundary of the object $x=\frac{2}{d-3}$. By looking at~(\ref{zepre}), we see this solution trivially satisfies the condition $(d-2)p_{\bot}\leq \rho$.

\section{Discussion}

In this work we have investigated anisotropic spherically symmetric matter models in higher dimensions. We have derived bounds on the ratio of the gravitational mass $m_g(r)$ to the radius $r$. Various energy conditions are considered, and the effects of both a positive cosmological constant and non-zero charge are also examined. These bounds generalise inequalities found by previous authors~\cite{Andreasson:2007ck,Andreasson:2008xw,Andreasson:2009pe,Andreasson:2012dj,Bondi:1964zz,Karageorgis:2007cy} to dimensions greater than four. A new inequality in four dimensions has also derived~(\ref{dominant2}) when the energy density is greater than the tangential pressure for the case of positive $\Lambda$ and non-zero charge.

We have examined three distinct matter models; assuming the radial pressure is greater than the tangential pressure, assuming the energy density is greater than the tangential pressure, and assuming the Andr\`{e}asson condition
\begin{align}
p+(d-2)p_{\bot}\leq \Omega\rho.
\end{align} 
These energy conditions incorporate a wide class of physically meaningful situations including any matter with positive energy and pressure obeying the dominant energy condition,  and infinitely thin bubbles with zero radial pressure. 

The inequalities derived allow us to bound the maximum gravitational redshift, and examine which dimension the result is greatest. In all our bounds, our results confirm the conclusion that the effects of gravity are weaker in higher dimensions.  We see that assuming the Andr\`{e}asson condition with $\Omega\leq 1$ gives us the strictest bound on the mass-radius ratio and gravitational redshift in higher dimensions. We also recover the original Buchdahl bound in higher dimensions~(\ref{Buchdahld}) by setting $\Omega=1$ without the need for assuming Buchdahl's restrictive assumptions. All of the bounds obtained agree with the conclusion that the maximum mass radius ratio occurs before the coordinate horizon located at 
\begin{align}
R=\left(\frac{2\kappa M}{(d-2)A_{d-2}}\right)^{\frac{1}{d-3}},
\end{align}
and thus a regular matter distribution cannot describe the interior of a higher dimensional black hole.

\begin{center}
\textbf{Acknowledgement}
\end{center}
The author would like to thank Christian B\"{o}hmer for useful discussions and comments on the manuscript. We also wish to thank the anonymous referees for helpful feedback and suggestions on improving the manuscript.

\end{document}